\def\be{\begin{equation}}
\def\ee{\end{equation}}
\def\bea{\begin{eqnarray}}
\def\eea{\end{eqnarray}}
\def\av#1{\langle {#1} \rangle}
\def\r{{\bf{r}}}
\def\ro{{\bf{r_o}}}
\documentstyle[prl,epsfig,aps,multicol]{revtex}
\begin{document}
\draft
\title{Intensity distribution for waves in disordered media:
deviations from Rayleigh statistics}

\author{A.D. Mirlin$^a$, R. Pnini$^b$ and B. Shapiro$^c$}
\address{$^a$Institute f{\"u}r Theorie der Kondensirerten Materie, 
Universit{\"a}t Karlsruhe, 76128 Karlsruhe, Germany} 
\address{$^b$Department of Pure and Applied Sciences, University of Tokyo,
Komaba, Meguro-ku, Tokyo 153, Japan}
\address{$^c$Department of Physics, Technion - Israel Institute 
of Technology, 32000 Haifa, Israel}

\date{\today}       
\maketitle
\begin{abstract}
We study the intensity distribution function, $P(I)$,
for monochromatic waves propagating in quasi one-dimensional
disordered medium, assuming that a point source and a point detector
are embedded in the bulk of the medium.
We find deviations from the Rayleigh statistics at moderately large
$I$ and a logarithmically-normal asymptotic behavior of $P(I)$.
When the radiation source and the detector are located close to the
opposite edges of the sample (on a distance much less then the sample
length), an intermediate regime with a stretched-exponential behavior
of  $P(I)$ emerges. 
\end{abstract}

\begin{multicols}{2}

\noindent 1.  When a wave propagates through a 
random medium, it
undergoes multiple scattering from inhomogeneities.
The scattered intensity pattern (speckle pattern)
is highly irregular and should be described in 
statistical terms. One of its characteristics is 
the intensity distribution, $P(I)$, at some point
$\r$. Almost a century ago Lord Rayleigh, using
simple statistical arguments, proposed a distribution
which bears his name:
\be
P_o(\tilde{I})= \exp(-\tilde{I}) \ ,
\label{rayleigh}
\ee
where $\tilde{I}$ is the intensity normalized to its average
value, $\tilde{I}=I/\av{I}$. The Rayleigh distribution has moments
$\av{\tilde{I}^n}=n!$ and it provides, in many cases, a rather
accurate fit to experimental data, as long as $I$ is 
not too large \cite{ishimaru}. For large $I$, however,
the data show large deviations from Eq.~(\ref{rayleigh})
\cite{ishimaru,dainty,genack}. Various extensions of 
Eq.~(\ref{rayleigh}) have been proposed in the literature.
Jakeman and Pusey \cite{pusey} proposed to fit the data with
the K-distribution. It contains a phenomenological
parameter $\eta$ and its moments are given by 
$\av{\tilde{I}^n}=n!\eta^{-n}\Gamma(n+\eta )/\Gamma(\eta )$.
The experimentally relevant situation corresponds to 
$\eta\gg 1$. In this case moments up to $n\lesssim\eta$ 
can be approximated as
\be
\av{\tilde{I}^n}\simeq n!\exp (n^2/\eta) \ ,
\label{moments}
\ee
where only the leading term in the exponent has been
kept. Thus, only low moments $(n\ll\sqrt{\eta})$ are 
close to the Rayleigh value $n!$. 
Some theoretical support to the phenomenological formula
Eq.~(\ref{moments}) has been 
given by Dashen \cite{dashen} who considered smooth
disorder (typical size of inhomogeneities much larger
than the wavelength). 

More recently, there was a considerable theoretical activity in
studying the statistics of the transmission coefficients $t_{ab}$ of a
one-dimensional sample with short-range disorder
\cite{kk1,nieu,kk2,been}. In this formulation of the problem, a source
and a detector of the radiation are located outside the sample. The
source produces a plane wave injected in an incoming channel $a$, and
the intensity in an outgoing channel $b$ is measured. It was shown in
\cite{nieu,kk2,been} that the distribution of the normalized
transmission coefficients $s_{ab}=t_{ab}/\langle t_{ab}\rangle$
 crosses over
from the Rayleigh distribution $P(s_{ab})=e^{-s_{ab}}$ to a
stretched-exponential one $P(s_{ab})\sim e^{-2\sqrt{gs_{ab}}}$, where
$g$ is the dimensionless conductance.

In this paper we consider a different situation, when both the source
and the detector of the radiation are embedded into the bulk of the
sample, and calculate the intensity distribution $P(I)$ in this case.
We prove that, for not too large $n$, the moments can be indeed
described by Eq.(\ref{moments}), and compute the parameter $\eta$
phenomenologically introduced in \cite{pusey}. We compute further the
whole distribution function $P(I)$ and show that its asymptotic
behavior at large $I$ is of a logarithmically-normal form, in contrast
to the stretched-exponential asymptotics of $P(t_{ab})$ found in
Refs.\cite{nieu,kk2,been}. Finally, we discuss how these two different
forms of the asymptotic behavior match each other and describe
physical mechanisms governing both of them.

\begin{figure}
\narrowtext
\centerline{\epsfig{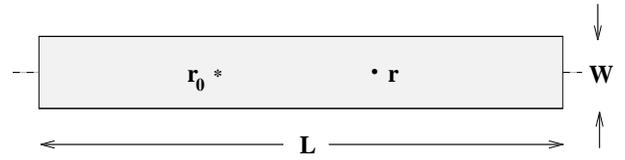}}
\vspace{0.3cm}
\caption{Geometry of the problem.
Points $\ro=(x_o,y_o,z_o)$ and
$\r=(x,y,z)$ are the positions of the source
and of the observation point respectively.}
\label{tube}
\end{figure}

\begin{figure}
\narrowtext
\centerline{\epsfig{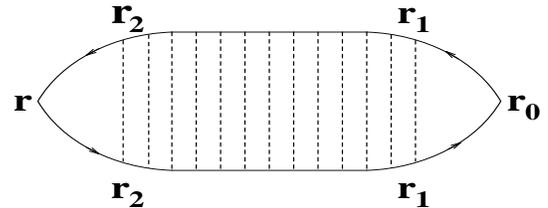}}
\vspace{0.3cm}
\caption{Diagram for the average intensity. The diffusion
ladder is inserted between two solid lines which represent
the average Green's functions.}
\label{ladder}
\end{figure}

\noindent 2.  We assume a quasi one-dimensional 
geometry, i.e.\ consider a tube of transverse dimension 
$W$ and length $L\gg W$, filled with scattering medium
(Fig.~\ref{tube}). The monochromatic source of radiation
is placed at point $\ro$. The field at some point
$\r$ is given by the (retarded) Green's function
$G_R (\r,\ro)$ and the radiation intensity
is defined as $I(\r,\ro)
\equiv |G_R (\r,\ro)|^2$.
The average intensity $\av{I(\r,\ro)}$
is represented diagrammatically in Fig~\ref{ladder}.
It consists of a diffusion ladder (a diffuson),
$T({\bf{r_1}},{\bf{r_2}})$, attached to two external
vertices. The vertices are short-range objects and can 
be approximated by a $\delta$-function times 
$(\ell/4\pi )$, so that
$\av{I(\r,\ro)}=
(\ell/4\pi )^2 T({\bf{r}},\ro)$.
For the quasi one-dimensional geometry, the expression for  the
diffuson reads 
\be
T({\bf r},{\bf r_o})=\left(\frac{4\pi}{\ell}\right)^2
\frac{3}{4\pi}\frac{\left[z_{<}(L-z_{>})\right]}
{A\ell L}
\label{diffuson}
\ee
where $\ell$ is the elastic mean free path, $A$
is the cross-section of the tube, 
$z$-axis is directed  along the sample,
$z_{<}=\text{min} (z,z_o)$ 
and $z_{>}=\text{max} (z,z_o)$. We assume, of course, that 
$|z-z_0|\gg W$. 

The intensity distribution $P(I)$, in the diagrammatic
approach,  is obtained by calculating the moments
$\av{I^n}$ of the intensity. In the leading approximation
\cite{boris}, one should draw $n$ retarded and $n$ advanced
Green's functions and insert ladders between pairs
$\left\{G_R,G_A\right\}$ in all possible ways. This leads
to $\av{I^n}=n!\av{I}^n$ and, thus, to Eq.~(\ref{rayleigh}).

\begin{figure}
\narrowtext
\centerline{\epsfig{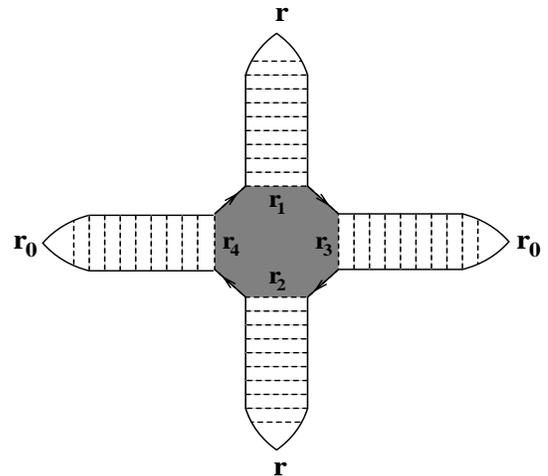}}
\caption{Diagram for a pair of interacting diffusons. 
The external vertices contribute the factor $(\ell/4\pi)^4$.
The shaded region denotes the internal interaction vertex, see Eq.~\ref{C2}}
\label{hbox}
\end{figure}

Corrections to the Rayleigh result come from diagrams with
intersecting ladders, which describe interaction between
diffusons. The leading correction is due to
pairwise interactions. The diagram in Fig.~\ref{hbox} 
represents a pair of "colliding" diffusons. The algebraic 
expression for this diagram is 
\bea
 C(\r,\ro) &=& 2 \left(\frac{\ell}{4\pi}\right)^4
\int\left(\prod_{\imath=1}^4 d^3{\bf{r_\imath}}\right) \nonumber\\  
&\times & T(\r,{\bf{r_1}}) T(\r,{\bf{r_2}})
T({\bf{r_3}},\ro) T({\bf{r_4}},\ro)  \nonumber\\
&\times & \left\{
\left( \frac{\ell^5}{48\pi k^2_o} \right)
\int d^3{\bbox{\rho}} \left[ 
(\nabla_1 +\nabla_2)\cdot (\nabla_3 +\nabla_4)\right. \right.\nonumber
\\ 
& + & 2 \left. \left.
(\nabla_1\cdot\nabla_2) + 2 (\nabla_3 \cdot\nabla_4)\right]
\prod_{\imath=1}^4 \delta({\bbox{\rho}}-{\bf{r_\imath}})
\right\} \ ,
\label{C2}
\eea
where  $k_o$ is the wave number and 
$\nabla_{\imath}$ acts on ${\bf{r_\imath}}$.
The factor $\left(\ell/4\pi\right)^4$ comes from the 
4 external vertices of the diagram, the $T$'s represent
the two incoming and two outgoing diffusons and
the expression in the curly brackets corresponds to 
the internal (interaction) vertex \cite{hikami}.
Finally, the factor 2 accounts for the two possibilities
of inserting a pair of ladders between the outgoing
Green's functions. Integrating by parts and employing
the quasi one-dimensional geometry of the problem, we obtain
(for $z_o<z$):
\be
 C(z,z_o) \simeq 2 \av{I(z,z_o)}^2
\left( 1 + \frac{4}{3\gamma}\right) \ ,
\label{Cgamma}
\ee
where $\av{I(z,z_o)}={3\over 4\pi}{z_0(L-z)\over A\ell L} $ is the average
intensity, 
\be
\gamma = 2g {L^3\over L^2(3z+z_o) - 2Lz(z+z_o) + 2z^2_o (z-z_o)} \gg 1
\ ,
\label{gamma}
\ee
and $g=k^2_o\ell A/3\pi L\gg 1$ is the dimensionless
conductance of the tube.
For simplicity, we will assume that the source and the detector are
located relatively close to each other, so that  $|z-z_o|\ll L$, in
which case Eq.(\ref{gamma}) reduces to $\gamma= gL^2/2z(L-z)$. (All the
results are found to be qualitatively the same in the generic
situation $z_0\sim z-z_0 \sim L-z\sim L$.)

In order to calculate $\av{I^n}$ one has to
compute a combinatorial factor which counts
the number $N_i$ of diagrams with $i$ pairs of
interacting diffusons. This number is \cite{kk1}
$N_i=(n!)^2/[{2^{2i}}{i!}{(n-2i)!}]\simeq
(n!/i!)(n/2)^{2i}$,
so that
\be
\frac{\av{I^n}}{\av{I}^n}=
n!\sum_{\imath=0}^{[n/2]} \frac{1}{i!}
\left(\frac{2n^2}{3\gamma}\right)^i
\simeq n!\exp(2n^2/3\gamma)  \  .
\label{moments2}
\ee
Although $i$ cannot exceed $n/2$, the sum in
Eq.~(\ref{moments2}) can be extended to $\infty$, if the
value of $n$ is restricted by the condition $n\ll\gamma$. 
Eq.~(\ref{moments2}) represents the leading exponential correction to
the Rayleigh distribution. Let us discuss now the effect of higher order
``interactions'' of diffusons. 
Diagrams with 3 intersecting diffusons will contribute
a correction of $n^3/\gamma^2$ in the exponent of
Eq.~(\ref{moments2}), which is small compared to the leading correction
in the whole region $n\ll\gamma$, but becomes larger than unity 
for $n\gtrsim\gamma^{2/3}$. Likewise, diagrams with 4 intersecting
diffusons produce a $n^4/\gamma^3$ correction, etc.
Restoring the distribution $P(I)$, we find
\be
P(\tilde{I})\simeq\exp\left(-\tilde{I}+\frac{2}{3\gamma}\tilde{I}^2
+O\left({\tilde{I}^3\over\gamma^2}\right)+\ldots\right) \ ,
\label{intermediate}
\ee
It should be realized that Eq.~(\ref{intermediate})
is applicable only for $\tilde{I}\ll \gamma\sim g$ and, thus, does not
determine the far asymptotics of $P(I)$. The latter is
unaccessible by the perturbative diagram technique and 
is handled below by the supersymmetry method.

\noindent 3.  In the supersymmetry formalism, averaging 
over disorder is replaced by functional integration over
supermatrix fields, $Q(\r)$, which satisfy the 
constraint $Q^2=1$ \cite{efetov}. For technical simplicity, we will
assume that the time reversal symmetry is broken by some
magnetooptical effects.
The integration is done with a weight function 
$\exp[-S\{Q\}]$, where $S\{Q\}$ is the \mbox{$\sigma$-model}
action, 
\be
S\{Q\}=-\frac{\pi\nu D}{4}\int d^3\r\, {\rm Str}
(\nabla Q)^2 \ ,
\label{sigma}
\ee 
Str denotes the supertrace, 
$D$ is the diffusion constant,
and $\nu$ is the average density of states. 
In the considered quasi-1D geometry,  
$\pi\nu D=gL/2A$, and the field $Q$ depends on the $z$-coordinate
only, yielding
$S\{Q\}=-(gL/8) \int dz\, {\rm Str}(dQ/dz)^2$.
Following the derivation outlined
in~\cite{mir1,efpri,mir3}, the moments of the 
intensity at point $\r$ due to the source at
$\ro$ are given by
\be
\av{I^n}= \left(-\frac{k_0^2}{16\pi^2}\right)^n 
\int [{\cal D}Q] 
\left(Q^{bb}_{12} (z)\right)^n
\left(Q^{bb}_{21} (z_0)\right)^n
e^{-S\{Q\}},
\label{superI}
\ee
where $Q^{bb}_{12}(Q^{bb}_{21})$ is
the retarded-advanced (resp. advanced-retarded) matrix 
element in the boson-boson sector of $Q$. 
Assuming again that the two points
$\r$ and $\ro$ are sufficiently close to each other, $|z-z_0|\ll L$
 and taking into account slow variation of the $Q$-field
along the sample, we can replace the product 
$Q^{bb}_{12}(z) Q^{bb}_{21}(z_0)$ 
by 
$Q^{bb}_{12}(z)Q^{bb}_{21}(z)$. 
We get then the following result for the distribution of the
dimensionless intensity $y= (16\pi^2/k_0^2)I$:
\be
P(y)=\int dQ \delta( y+ Q^{bb}_{12} Q^{bb}_{21})Y(Q) \ ,
\label{dist1}
\ee
where $Y(Q)$ is a function of a single supermatrix $Q$ defined as
follows \cite{mir1,mir3}: 
\be
Y(Q_o)\equiv \int_{Q(\ro)=Q_o} 
[{\cal D}Q] \exp[-S\{Q\}]   \ .
\label{Yfunction}
\ee
In general, the function $Y(Q)$ depends only on the 
parameters $1\le\lambda_1<\infty$, 
$-1\le\lambda_2\le 1$
entering the standard parametrization of the $Q$-matrices
\cite{zirnbauer}. 
Performing the integration over the other degrees of freedom, we find
\bea
P(y)&=&\left(\frac{d}{dy} +y\frac{d^2}{dy^2}\right)
\int d\lambda_1 d\lambda_2 \nonumber \\
&\times& \left(\frac{\lambda_1 +\lambda_2}{\lambda_1 -\lambda_2}
\right) Y(\lambda_1 ,\lambda_2 )
\delta(y + 1-\lambda^2_1) \ .
\label{dist2}
\eea
The evaluation of $Y(Q_o)=Y(\lambda^o_1 ,\lambda^o_2)$ 
involves, by its definition~(\ref{Yfunction}),  
an integration 
over all supermatrix fields, 
which assume a given value $Q_o$ 
at point $z_0$ and satisfy the 
boundary conditions $Q|_{z=0,L}=\Lambda$, where
$\Lambda\equiv{\rm diag}\{1,1,-1,-1\}$.
Since  $g\gg 1$, this calculation
can be done by the saddle point method, as
suggested by Muzykantskii and Khmelnitskii \cite{musyk}. 
The result is \cite{mir3}
\be
Y(\lambda_1 ,\lambda_2) 
\simeq\exp\left\{-\frac{\gamma}{2}
\left[\theta^{2}_1 +\theta^{2}_2 \right]
\right\} \ .
\label{SP}
\ee
where $\lambda_1\equiv\cosh\theta_1$, $\lambda_2\equiv\cos\theta_2$
($0\leq\theta_1<\infty$, $0\leq\theta_2\leq\pi$).
In fact, the dependence of $Y$ on $\theta_2$ is
not important, within the exponential accuracy,
because it simply gives a prefactor after the
integration in Eq.~(\ref{dist2}). Therefore, up to  a 
pre-exponential factor, 
the distribution function $P(y)$ 
is given by 
\be
P(y)\sim Y(\lambda_1=\sqrt{1+y} ,\lambda_2=1)
\sim
\exp\left(-\gamma\theta^{2}_1 /2 \right) \ , 
\label{dist3}
\ee 
where $\theta_1 =\ln (\sqrt{1+y}+\sqrt{y})$.
Finally, after normalizing $y$ to its average value
$\av{y}=2/\gamma$, we obtain: 
\be
P(\tilde{I})\simeq\exp\left\{-\frac{\gamma}{2} 
\left[\ln^2
\left(\sqrt{1+2\tilde{I}/\gamma} +\sqrt{2\tilde{I}/\gamma}\right)
\right]\right\} \ .
\label{final}
\ee
For $\tilde{I}\ll \gamma$, Eq.(\ref{final}) reproduces the
perturbative expansion (\ref{intermediate}), while for
$\tilde{I}\gg\gamma$ it implies the log-normal asymptotic behavior
of the distribution $P(\tilde{I})$:
\be
\ln P(\tilde{I}) \simeq -(\gamma/8)\ln^2(8\tilde{I}/\gamma)\ .
\label{LN}
\ee

\noindent 4. The log-normal ``tail'' (\ref{LN}) should be contrasted
with the stretched-exponential  asymptotic behavior of the distribution
of transmission coefficients \cite{nieu,kk2,been}. Let us briefly
discuss, how these two results match each other. Analyzing the
expression for the moments (\ref{superI}), we find that when the
points $z$ and $z_0$ approach the sample edges, $z_0=L-z\ll L$, an
intermediate regime of stretched-exponential behavior emerges:
\begin{equation}
\ln P(\tilde{I})\simeq\left\{
\begin{array}{ll}
-\tilde{I}+{1\over 3g}\tilde{I}^2+\ldots\ , & \ \ \tilde{I}\ll g \\
-2\sqrt{g\tilde{I}}\ , & \ \ g\ll \tilde{I}\ll g\left({L\over
z_0}\right)^2\\ 
-{gL\over 8z_0}\ln^2\left[16\left({z_0\over L}\right)^2
{\tilde{I}\over g}\right]\ , & \ \
\tilde{I}\gg g\left({L\over z_0}\right)^2 \ .
\label{strexp}
\end{array} \right.
\end{equation}
Thus, when the source and the detector move toward the sample edges,
the region of validity of the stretched-exponential behavior becomes
broader, while the log-normal ``tail'' gets pushed further away. In
contrast, when the source and the detector are located deep in the
bulk, $z_0\sim L-z\sim L$, the stretched-exponential regime
disappears, and the Rayleigh distribution crosses over directly to the
log-normal one at $\tilde{I}\sim g$. 

Let us now describe the physical mechanisms standing behind these
different forms of $P(\tilde{I})$. The Green's function
$G^R({\bf r_0},{\bf r})$ can be expanded in eigenfunctions of a
non-Hermitean (due to open boundaries) ``Hamiltonian'' as 
$G^R({\bf r_0},{\bf r})=\sum_i\psi_i^*({\bf r_0})\psi_i({\bf r})
(k_0^2-E_i+i\gamma_i)^{-1}$. Since the level widths $\gamma_i$ are
typically of order of the Thouless energy $E_c\sim D/L^2$, there is
typically $\sim g$ levels contributing appreciably to the sum. 
In view of the random phases of the wave functions, this  
leads to a Gaussian
distribution of $G^R({\bf r_0},{\bf r})$ with zero mean, and thus
to the Rayleigh distribution of $I({\bf r_0},{\bf r})=|G^R({\bf
r_0},{\bf r})|^2$, with the moments $\langle\tilde{I}^n\rangle=n!$. 
The stretched-exponential behavior results from the disorder
realizations, where one of the states $\psi_i$ has large amplitudes in
the both points ${\bf r_0}$ and ${\bf r}$. Considering both 
$\psi_i({\bf r_0})$ and $\psi_i({\bf r})$ as independent random
variables with Gaussian distribution and taking into account
that only one (out of $g$) term contributes in this case
to the sum for $G^R$, we find
$\langle\tilde{I}^n\rangle\sim n!n!/g^n$, corresponding to the above
stretched-exponential form of $P(\tilde{I})$. Finally, the log-normal
asymptotic behavior corresponds to those disorder realizations, 
where $G^R$ is dominated by an anomalously localized state, which has
an atypically small width $\gamma_i$ (the same mechanism determines
the log-normal asymptotics of the distribution of local density of
states, see Refs.~\cite{mir3,als}).

\acknowledgments 
A.D.M. gratefully acknowledges kind hospitality extended to him in the
Physics Department of the Technion, where most of this work was done,
and financial support from SFB195 der Deutschen Forschungsgemeinschaft.
This research was supported in part by a grant from the
Israel Science Foundation and by the Fund for promotion
of research at the Technion.

\end{multicols}
\end{document}